\documentstyle[preprint,prb,aps]{revtex}
\tightenlines
\begin{document}
\draft

\title{
Many-Polaron System Confined to a Quantum Dot:
Ground-State Energy and Optical Absorption
}
\author{J. T. Devreese$^{*}$, S. N. Klimin$^{**}$,
V. M. Fomin$^{**}$, and F. Brosens}

\address{Universiteit Antwerpen (U.I.A.), Departement Natuurkunde,\\
Universiteitsplein 1, B-2610 Antwerpen-Wilrijk, Belgi\"e}
\date{\today}
\maketitle

\begin{abstract}
\noindent
We find for the first time the ground state energy and the
optical absorption spectra for $N$ electrons (holes) interacting with
each other and with the longitudinal optical (LO) phonons at an arbitrary
electron-phonon coupling strength $\alpha$ in a parabolic confinement
potential. A recently developed path integral formalism for identical particles is
used in order to account for the fermion statistics. The approach is
applicable to closed and open shells.
Using an extension of the Jensen-Feynman variational principle,
the ground state energy of the $N$-polarons system in a parabolic confinement
potential is analyzed as a function of $N$ and $\alpha$.
A ferromagnetic-to-nonmagnetic transition is shown to occur between states
with different total spin of the system in the case of strong
electron-phonon interaction. This transition is manifested through the
optical absorption spectra and should be experimentally observable.
Strong mixing between zero-phonon and one-phonon states is
revealed in the optical absorption spectrum, when the confinement frequency
parameter is in resonance with the LO phonon frequency (``confinement-phonon
resonance'').
Moments of the optical absorption spectra are calculated for a $N$-polaron
system in a parabolic quantum dot.

\end{abstract}
\bigskip

%\pacs{PACS numbers: }
\narrowtext

\section{Introduction}

The path integral method for indistinguishable particles developed in Refs.
\onlinecite{PRE96,PRE97} seems an adequate tool for the investigation of both the
equilibrium and the non-equilibrium properties of interacting quantum
many-body
systems. The advantages of the path integral representation are particularly
manifest when investigating systems with a fixed (few or many) number of
particles. As demonstrated in Refs. \onlinecite{PRE96,PRE97}, the thermodynamical
properties of systems with a limited number of particles might deviate
substantially from those obtained in the thermodynamical limit.

Electrons confined to a quantum dot can provide a typical example of a
system with a fixed number of indistinguishable particles. In recent years,
the quantum states and the optical properties of these systems have received
considerable attention. Multi-electron states in 3D and 2D quantum dots
(without the electron-phonon interaction) have, e.~g., been treated in Refs.
\onlinecite{Maks,Taut,Gonz,Asho,Ryan,Fuji,Tar,Szafran,Hoh}. It has been shown in
Ref. \onlinecite{Hoh} that the interaction between the charge carriers strongly
influences the optical spectra of few-particle quantum dots. The bipolaron
ground state in quantum dots has been studied in Ref. \onlinecite{JP99}.
Cooperative effects for a gas of polarons and bipolarons were treated in
Ref. \onlinecite{Jad} on the basis of a model introduced by Friedberg and Lee
\cite{FL}. To the best of our knowledge, the effects of the
electron-phonon interaction on the optical spectra of multi-electron quantum
dots with a fixed number of electrons have hitherto not been investigated
theoretically.

Experimentally observed optical absorption spectra of high-$T_{c}$ cuprates
\cite{Genzel,Thomas,Calvani} are very promising to reveal new manifestations
of the interaction between electrons (holes) and the longitudinal
optical (LO) phonons.
A possible role of polarons in high-T$_c$ superconductivity has been analyzed
by several authors (see, e. g., Refs. \onlinecite{Alran,Emin,Rann,Alex}).
Some aspects of the recently observed optical
absorption spectra of high-$T_{c}$ superconductors \cite{Tsvetkov,Wang,Wang2}
can be interpreted using the polaron theory \cite{DSG,Green,PD83}. It has
been shown in Ref. \onlinecite{DT98} that certain characteristics of the polaron
optical absorption at intermediate Fr\"{o}hlich coupling constant $\alpha $
appear in the aforementioned spectra \cite{Genzel,Thomas,Calvani}. Bipolaron
optical absorption in bulk materials has, e.~g., been treated in Refs. \onlinecite
{Salje,DF96}. In view of the relatively high concentration of electrons
(holes) in high-$T_{c}$ superconductors, the development of an all-coupling
and all-concentration theory of the optical response of many-polaron systems
is an urgent problem.

\section{Ground-state energy}

In the present communication, we treat the ground state energy and the
optical absorption spectra of a multi-electron (multi-polaron) parabolic
quantum dot for arbitrary electron-phonon coupling strength, using a recently
developed path
integral formalism for the quantum statistical treatment of identical
particles \cite{PRE96,PRE97}
and considering the expected quite large polaron coupling constant
in those materials.
We consider a system consisting of $N$ electrons in a parabolic confinement
potential characterized by the frequency parameter $\Omega _{0}.$ These
electrons are assumed to interact with each other and with the LO phonons.
The electron subsystem is subdivided into two groups ($N=\sum_{\sigma
}N_{\sigma },$ where $N_{\sigma }$ is the number of electrons with definite
spin projection $\sigma =\pm 1/2$).

In the present work the ground state energy of $N$ polarons confined
to a quantum dot has been determined for the first time. This calculation
has been performed within the \emph{extended Jensen-Feynman variational principle}
\cite{PRE96} taking into account the symmetry properties of the
electrons with respect to permutations. The validity of this
extension of the Jensen-Feynman inequality for systems of indistinguishable
particles, which is not obvious, and which is of crucial importance for the
present work, has been demonstrated in Ref. \onlinecite{PRE96}. We have
used an auxiliary model
system of particles in a harmonic confinement potential with elastic
interparticle interactions as studied in Refs. \onlinecite{PRE96,PRE97}.
The parameters of the auxiliary model system have been treated as variational
parameters. We have worked out the variational procedure to obtain
those variational parameters at arbitrary temperature for the physical system
of $N$ polarons confined to a parabolic potential. The calculation
has been performed for the case of closed shells (i.~e. of a non-degenerate
ground state) and for the case of open shells.
%The open-shell problem remains to be analyzed in detail.

We have calculated the ground state energy of a parabolically confined
many-polaron system as a function of the total spin of the system $S=\frac{1%
}{2}\left| N_{+1/2}-N_{-1/2}\right| $.
For relatively small polaron coupling constant and all values of
$\eta\equiv \varepsilon
_{\infty }/\varepsilon _{0}$ ($\varepsilon _{\infty }$ and $\varepsilon _{0}$%
\ are the high-frequency and the static dielectric constants, respectively),
except $\eta \ll 1$, the electrons tend to maximally fill up {\it open}
shells with the
\emph{same} spin projection (cf. the {\it first Hund's rule}).
In this domain of the parameters $\alpha$ and $\eta$, the ground state is
obtained when at least one of the numbers $N_{\pm 1/2}$ equals the number of
states ${M}_{n}$ in a set of shells labelled by the
index $k$ from 0 to $n$ (see Ref. \onlinecite{PRE97}):
\begin{equation}
{M}_{n}=
\sum_{k=0}^{n}g_{k}=\frac{\left( n+1\right) \left( n+2\right) \left(
n+3\right) }{6} \>\>\>(n=0,1,2,\dots),
\label{Magic}
\end{equation}
where $g_{k}\equiv \frac{1}{2}\left( k+1\right) \left( k+2\right) $ is the
degeneracy of the $k$th energy level of a three-dimensional oscillator.
For $n=0,1,2,3\dots$ the values of the number $M_n$ are $1,4,10,20\dots$.

This filling scheme
is \emph{broken} for strongly polar substances ($\eta \ll 1$), where we
find that the lowest variational energy corresponds to the minimal possible
total spin ($S=0$ for $N$ even, and $S=\frac{1}{2}$ for $N$ odd), or for
$\alpha \gg 1$ (except the region $\eta \ll 1$), where the lowest
variational energy corresponds to the maximal possible total spin ($S=N/2$).

%Such a
%behavior of the ground state energy indicates that the phonon-mediated
%attraction between the electrons in this case is able to overcome their
%Coulomb repulsion.

In Table 1, the variational ground state energy per particle $E_{0}/N$
is shown for $N=20$\ polarons in a quantum dot with confinement
frequency parameter $\Omega _{0}=0.5$\ at $\alpha =5$, for
different values of $\eta .$

\begin{center}
\textbf{Table 1.} The variational ground state energy per particle $E_{0}/N$
(in units of the LO phonon energy $\hbar \omega _{\mathrm{LO}}$)
for different
$\eta$  for $N=20$\ polarons in a quantum dot with the confinement
frequency parameter $\Omega _{0}=0.5$\ (in units of $\omega _{\mathrm{LO}})$\
at $\alpha =5.$

\medskip
\begin{tabular}{|c|c|c|}
\hline
$\eta $ & $E_{0}/N\left(
\begin{array}{c}
N_{+\frac{1}{2}}=20, \\
N_{-\frac{1}{2}}=0 \\  [0.1cm]
\end{array} 
\right) $ & $E_{0}/N\left(
\begin{array}{c}
N_{+\frac{1}{2}}=10, \\
N_{-\frac{1}{2}}=10  \\ [0.1cm]
\end{array}
\right) $  \\   
\hline
0.01 & $-3.\,9404$ & $-4.1078$ \\ \hline
0.081 & $-2.0867$ & $-2.0873$ \\ \hline
0.082 & $-2.0626$ & $-2.0620$ \\ \hline
0.4 & \phantom{$-$}$5.8487$ & \phantom{$-$}$6.0482$ \\ \hline
\end{tabular}
\end{center}

\noindent As follows from this table, a ferromagnetic-to-nonmagnetic
transition takes place at a value of $\eta$ in the interval between
0.081 and 0.082.\ It will be analyzed below
how this transition influences the optical absorption spectra.
%A detailed
%description of the ground state energy for a many-polaron system confined in
%a quantum dot will be presented elsewhere.

\section{Optical absorption spectra}

In order to investigate the optical properties of the confined
many-polaron system,
we use the formalism developed in Refs. \onlinecite{DSG,Green,PD83}. An
alternative derivation can be found in Ref. \onlinecite{DF96}. Within this
technique, the absorption spectrum for a many-polaron system in a parabolic
confinement potential is given by the expression
\begin{equation}
\Gamma \left( \omega \right) \sim \frac{-\mbox{Im}\chi \left( \omega \right) }{%
\left[ \omega -\Omega _{0}^{2}/\omega -\mbox{Re}\chi \left( \omega \right) %
\right] ^{2}+\left[ \mbox{Im}\chi \left( \omega \right) \right] ^{2}},
\label{Kw}
\end{equation}
where the memory function is
\begin{eqnarray}
\chi \left( \omega \right) =\sum_{\mathbf{q}}\frac{2\left| V_{\mathbf{q}%
}\right| ^{2}q^{2}}{3N\hbar \omega }\int\limits_{0}^{\infty }\left(
e^{i\omega t}-1\right) \mbox{Im}\left[ T_{\omega _{\mathrm{LO}}}^{\ast }\left( t\right)
\left\langle \rho _{\mathbf{q}}\left( t\right) \rho _{-\mathbf{q}}\left(
0\right) \right\rangle _{S_{M}}\right] \,dt.  \label{memfunc}
\end{eqnarray}
The function $T_{\omega }\left( t\right) =\cos \left[ \omega \left( t-i\beta
/2\right) \right] /\sinh \left( \beta \omega /2\right) $ with $\beta =\hbar
/k_{B}T$ describes the phonon dynamics. The time-dependent correlation
function $\left\langle \rho _{\mathbf{q}}\left( t\right) \rho _{-\mathbf{q}%
}\left( 0\right) \right\rangle _{S_{M}}$ is the path integral average for an
auxiliary model action functional $S_{M}$ of $N$ identical electrons and $N_{B}$
identical fictitious particles (simulating the influence of the
LO phonon bath). As emphasized above, the parameters of the auxiliary
model system are determined using the variational procedure for the
physical system of $N$ confined polarons under consideration.
Note that, formally, the term $\Omega
_{0}^{2}/\omega $ plays a role similar to the cyclotron frequency $\omega
_{c}$ in the theory of the cyclotron resonance of polarons \cite{CR}. In the
zero temperature limit, we have obtained the following analytic expression
for the memory function:
\begin{eqnarray}
&&\chi \left( \omega \right) =\frac{2\alpha m^{\ast }}{3\pi N\omega }\left(
\frac{\omega _{\mathrm{LO}}}{A}\right) ^{3/2}\sum_{p_{1}=0}^{\infty
}\sum_{p_{2}=0}^{\infty }\sum_{p_{3}=0}^{\infty }\frac{\left( -1\right)
^{p_{3}}}{p_{1}!p_{2}!p_{3}!}
%\nonumber  \\ \times
\left( \frac{a_{1}^{2}}{N\Omega _{1}A}\right) ^{p_{1}}\left( \frac{%
a_{2}^{2}}{N\Omega _{2}A}\right) ^{p_{2}}\left( \frac{1}{NwA}\right) ^{p_{3}}
\nonumber    \\
&\times& \left\{ \sum_{m=0}^{\infty }\sum_{n=0}^{\infty }\sum_{\sigma
}N_{m,\sigma }\left( 1-N_{n,\sigma }\right)
\right.
%\nonumber      \\ \times
\left[ \frac{1}{\omega -\omega _{\mathrm{LO}}-\left[ p_{1}\Omega
_{1}+p_{2}\Omega _{2}+\left( p_{3}-m+n\right) w\right] +\mathrm{i}%
\varepsilon }\right. \nonumber      \\
&-&\frac{1}{\omega +\omega _{\mathrm{LO}}+p_{1}\Omega _{1}+p_{2}\Omega
_{2}+\left( p_{3}-m+n\right) w+\mathrm{i}\varepsilon }
%\nonumber    \\
\left. +2\frac{\mathcal{P}}{\omega _{\mathrm{LO}}+p_{1}\Omega
_{1}+p_{2}\Omega _{2}+\left( p_{3}-m+n\right) w}\right]   \nonumber    \\
&\times&\sum_{l=0}^{m}\sum_{k=n-m+l}^{n}\frac{\left( -1\right)
^{n-m+l+k}\Gamma \left( p_{1}+p_{2}+p_{3}+k+l+\frac{3}{2}\right) }{\left(
k\right) !\left( l\right) !} \nonumber     \\
&\times& \left( \frac{1}{wA}\right) ^{l+k}
\left(\begin{array}{c}{n+2}\\{n-k}\end{array}\right)
\left(\begin{array}{c}{2k}\\{k-l-n+m}\end{array}\right)
%\nonumber  \\
+\left[ \frac{1}{\omega -\omega _{\mathrm{LO}}-\left( p_{1}\Omega
_{1}+p_{2}\Omega _{2}+p_{3}w\right) +\mathrm{i}\varepsilon }\right.
\nonumber  \\
&-&\frac{1}{\omega +\omega _{\mathrm{LO}}+p_{1}\Omega _{1}+p_{2}\Omega
_{2}+p_{3}w+\mathrm{i}\varepsilon }
%\nonumber     \\
\left. +2\frac{\mathcal{P}}{\omega _{\mathrm{LO}}+p_{1}\Omega
_{1}+p_{2}\Omega _{2}+p_{3}w}\right] \sum_{m=0}^{\infty }\sum_{n=0}^{\infty
}\sum_{\sigma ,\sigma ^{\prime }}N_{m,\sigma }N_{n,\sigma ^{\prime }}
\nonumber  \\
&\times&\sum_{k=0}^{n}\sum_{l=0}^{m}\frac{\left( -1\right) ^{k+l}\Gamma
\left( p_{1}+p_{2}+p_{3}+k+l+\frac{3}{2}\right) }{k!l!}
%\nonumber    \\
\left. \left( \frac{1}{wA}\right) ^{k+l}
\left(\begin{array}{c}{n+2}\\{n-k}\end{array}\right)
\left(\begin{array}{c}{m+2}\\{m-l}\end{array}\right)
\right\} ,
\end{eqnarray}
where $\varepsilon \rightarrow +0$,
$\mathcal{P}$ denotes the principal value, $A\equiv \left[
\sum_{i=1}^{2}a_{i}^{2}/\Omega _{i}+\left( N-1\right) /w\right] /N.$
$\Omega _{1},$ $\Omega _{2}$
and $w$ are the eigenfrequencies of the model system ($\Omega _{1}$\ is the
frequency of the relative motion of the center of mass of the electrons with
respect to the center of mass of the fictitious particles; $\Omega _{2}$\ is the
frequency related to the center of mass of the entire model system; $w$\ is
the frequency of the internal degrees of freedom), $%
a_{1}$ and $a_{2}$ are the coefficients of a canonical transformation which
diagonalizes the model Lagrangian, and $N_{n,\sigma }$ is the number of
electrons with spin projection $\sigma $ in the $n$th single-particle level
(shell).

\section{Discussion of results}

In Fig.~1, the optical absorption spectra of a many-polaron system
in a quantum dot
with parabolic confinement potential are plotted for $\Omega _{0}=0.5$
(all frequencies are measured in units of $\omega _{\mathrm{LO}}$), $\alpha
=1$ and for $N=5,10,14$ (panels $a, b, c,$ respectively). Due to the
confinement in all three dimensions,
the electron motion is fully quantized. Hence, when a photon is
absorbed, the electron recoil can be transferred only by discrete quanta. As
a result, the absorption spectrum consists of a series of $\delta $-like
peaks as distinct from the absorption spectrum of a bulk polaron. In this
and subsequent figures, the height  of each peak represents its intensity.

Fig.~2 shows, for reference, the optical absorption spectrum for
a \emph{single} polaron confined to
a quantum dot with $\Omega _{0}=0.2$ for $\alpha =5,$ and reveals several
essential elements. Following the nomenclature of
Refs. \onlinecite{DSG,Green} we
distinguish a central (zero-phonon) peak, peaks due to transitions to the
relaxed excited state (RES), and peaks associated with transitions to the
Franck-Condon state (FC). Also it is worth mentioning that there is the
``one-phonon shoulder'' in the optical absorption spectrum
with threshold at $\omega \sim 1$.
%The dashed curve represents the absorption spectrum of
%the bulk polaron at $\alpha =5$ \cite{DSG}.
The envelope of the absorption
spectrum for a polaron in a quantum dot is very similar to the bulk polaron
absorption spectra for the same $\alpha $, as first obtained in
Ref. \onlinecite{DSG}.
It follows that the obtained optical absorption
spectra for a polaron confined to a quantum dot are consistent with those
for a bulk polaron \cite{DSG,Green}.

As follows from Figs. 1 and 2, the $\delta $-like ``central peak'' for a
system of polarons, confined to a quantum dot with parabolic confinement
potential, is positioned at $\omega \sim \Omega_0.$
Without the electron-phonon interaction, the ``central peak'' for a
system of electrons,
confined to a quantum dot with parabolic confinement, is exactly at
the confinement parameter: $\omega = \Omega_0.$

Fig. 3 shows the evolution of the absorption spectrum as a function of the
confinement parameter $\Omega _{0},$ for $\Omega _{0}=0.5,$
$\Omega _{0}=0.8,$ $\Omega _{0}=1$
and $\Omega _{0}=1.2$. This calculation is performed for $N=4,$ $\alpha =3,$
$\eta =0.3.$ Near $\Omega _{0}=1,$\ the zero-phonon and the one-phonon peaks
are of comparable oscillator strength. A formal analogy (particularly
manifest for sufficiently small $\alpha )$\ exists between the resonance
condition $\Omega _{0}=1$\ discussed here and the wee-established
magnetophonon resonance around $\omega_{c}=1$
(Refs. \onlinecite{GF,Barnes}). We suggest the designation
``confinement-phonon resonance''.

Fig.~4 illustrates the ferromagnetic-to-nonmagnetic transition induced by
the increase of the electron-phonon interaction for a
many-polaron system confined to a quantum dot. In this figure, optical absorption spectra are
plotted for $N=20$ polarons in a quantum dot with confinement parameter $%
\Omega _{0}=0.5,$ and with electron-phonon coupling constant $\alpha =5.$
The parameter $\eta $ varies from $\eta =0.01$ to $\eta =0.4,$ and we find
that, as a consequence, the total spin of the ground state jumps from $S=0$
to $S=10$. As seen from Table 1, for $\eta \leq 0.081,$ the ground state
with $S=0$ is energetically favorable. The optical absorption spectra for
this case are shown on panels \emph{a }($\eta =0.01$) and \emph{b} ($\eta
=0.081$) of Fig.~4. A series of pronounced peaks clearly related to the
internal polaron excitations (cf. RES of a single polaron in Fig. 2) is seen
in the spectrum for the case of a strongly polar substance ($\eta =0.01$).
With increasing $\eta ,$ the relative intensity of these ``RES''
(many-polaron) peaks with respect to that of the zero-phonon peak decreases
as is seen from Fig.~4. This effect is due to weakening of the
electron-phonon interaction with increasing parameter $\eta.$

In the specific case under consideration ($N=20$),
when $\eta $ varies from $\eta =0.081$ to $\eta =0.082$, the total spin
abruptly changes from $S=0$ [$N_{1/2}=10$, $N_{-1/2}=10$]
to its maximal value $S=10$ [$N_{1/2}=20$ (0), $%
N_{-1/2}=0$ (20)], i.~e., the electrons become fully spin-polarized. Such a
jump in the magnitude of the total spin can also be realized by varying the
confinement parameter at fixed $\alpha $ and $\eta .$
%for all values of
%$\eta$ except $\eta\ll 1$, the number of electrons (polarons) with parallel
%spin in the ground state
%takes the values ${M}_{n}$ according  to Eq. (\ref{Magic}). ,
%occurs to be ${M}_{n}=10$ or 20.
Comparing the optical absorption spectra in panels \emph{b} and \emph{c} of
Fig.~4, one observes that the optical absorption spectrum as a whole shows an
abrupt change at the ferromagnetic-to-nonmagnetic transition, though
the shift of the most intense phonon sidebands towards lower frequencies
(with increasing $\eta $) continuously proceeds when $\eta$ passes through
the value 0.081. Note also the shift to lower frequencies of the most intense
phonon sidebands (again related to the ``many-polaron RES'') in Fig.~4$b$
compared to Fig.~4$a$.

\section{Moments of the optical absorption spectra}

We have
calculated (both for closed-shell and open-shell systems) the first frequency
moment $M_{1}$ of the optical absorption spectrum for a $N$-polaron system
confined to a quantum dot with parabolic confinement potential
\begin{eqnarray}
\left\langle \omega \right\rangle \equiv {{\mu_{1}}\over{\mu_0}}
= \frac{\int_{0}^{\infty }\omega
\Gamma \left( \omega \right) d\omega }{\int_{0}^{\infty }\Gamma \left(
\omega \right) d\omega }
\label{mom1}\end{eqnarray}
with $\Gamma \left( \omega \right) $ the optical absorption coefficient,
and the parameter
$\sigma \equiv \sqrt{\left\langle \left( \omega -\left\langle
\omega \right\rangle \right)^{2} \right\rangle}$, where
\begin{eqnarray}
\left\langle \left( \omega -\left\langle
\omega \right\rangle \right) ^{2}\right\rangle =
\left\langle \omega ^{2}\right\rangle -\left\langle \omega
\right\rangle ^{2}
\label{mom2}\end{eqnarray}
is the normalized second frequency moment of the optical absorption spectrum.

%Note that the $\delta $-like ``central peak'' for a system of polarons,
%{\em confined to a quantum dot} with parabolic confinement potential,
%is positioned at $\omega \sim \Omega_0.$ The
%``central peak'', along with peaks of the optical absorption spectrum, has
%been taken into account in the integrals $\int_{0}^{\infty }\omega \Gamma
%\left( \omega \right) d\omega ,$ $\int_{0}^{\infty }\Gamma \left( \omega
%\right) d\omega $.

The results for the first frequency moment of the optical ansorption spectrua
for $N$ polarons in a quantum dot with parabolic confinement
are shown in Fig.~5$a$. The first frequency moment is
plotted as a function of the effective density $N/{\cal V}$, where
${\cal V}=\left( 4\pi/3\right)\left(\hbar/
2m\Omega_{0}\right) ^{3/2}.$ The
points corresponding to definite $N$ are shown by filled circles.
The number of particles ranges from $N=1$ to $N=20.$ The points indicated
by arrows are related to the closed-shell systems.

The function $\langle \omega \rangle \left( N/{\cal V}\right)$
turns out to be non-monotonous. There is a maximum of the first frequency
moment $\langle \omega \rangle$ at $N=2$.
The first frequency moment has a minimum when the number of particles
takes the value $N=14$ corresponding to the closed-shell system ($N_{+1/2}=10,$
$N_{-1/2}=4$).

By its general trend, the first frequency
moment of the optical absorption spectrum as
a function of the effective density strikingly resembles
the first moment of optical conductivity spectrum in
Nd$_{2-\mbox{\small x}}$Ce$_{\mbox{\small x}}$CuO$_{4-{\rm y}}$
recently observed experimentally (Fig.~5$b$ and Ref. \onlinecite{cap99}).
Note that our theory describes a 3D confined system, while the experiment
relates to a quasi-2D
translationally invariant system. Therefore, the theory can be expected to
reveal only a {\it qualitative trend} of the first frequency
moment as a function of concentration in comparison with experiment.

\section{Conclusions}

We have presented the ground state energy and the optical absorption spectra
calculated for a system of $N$ polarons in a parabolic
confinement potential for any strength of the polaron coupling. Path
integral formalism for the quantum statistical physics of
indistinguishable particles \cite{PRE96,PRE97} has
allowed to develop the variational procedure \cite{PRE96} for the ground
state energy for a finite number of polarons.
For the first time, the ground state energy and the
optical absorption spectra have been analyzed
for $N$ electrons (holes) interacting with
each other and with the longitudinal optical (LO) phonons at an arbitrary
electron-phonon coupling strength $\alpha$ to a parabolic confinement
potential. A new type of transition for $N$ polarons confined in a parabolic potential (ferromagnetic-to-nonmagnetic
transition) is found between states with different total spin, which is
related to the competition between the \emph{Coulomb repulsion} and the
\emph{phonon-mediated attraction} between the electrons.
For relatively weak polaron
coupling constant, the electrons are shown to maximally fill up shells with
the same spin projection (cf. the first Hund's rule).
This filling scheme
is demonstrated to be broken for strongly polar substances ($\eta \ll 1$),
where we find that the lowest variational energy corresponds to the minimal
possible total spin.
%At present,
The present
analysis has been executed for closed-shell and open-shell systems.
%, but a more detailed
%investigation including open-shell systems is in progress.

The optical absorption spectra have been calculated here using the memory
function approach as applied to path integrals
for a many-polaron system confined to a quantum dot
with different number of polarons. The dependence of the optical absorption spectra on the
confinement parameter $\Omega _{0}$ reveals a resonant behavior for $\Omega
_{0}\approx 1$, especially if $\alpha $\ is small. The polaron
RES \cite{DSG,Green}
are seen to also influence the optical absorption spectra of $N$\ confined
polarons.

We have analysed also the first frequency moment of the optical absorption
spectrum for a $N$-polaron system in a parabolic quantum dot for both
closed-shell and open-shell systems.

\acknowledgements

This work has been supported by the BOF NOI (UA-UIA), GOA BOF UA 2000, IUAP,
FWO-V. projects G.0287.95, 9.0193.97, and the W.O.G. WO.025.99N (Belgium).
We are indebted to P. Calvani for fruitful interactions and communication of
experimental data. Discussions with L. F. Lemmens on the many body aspects
and a discussion with K. H. Michel on phase transitions are gratefully
acknowledged.

\end{document}